\documentclass[aps,prb,twocolumn,showpacs,amsmath,amssymb,preprintnumbers,article]{revtex4}

\usepackage{color}
\usepackage{graphicx}
\begin {document}

\title{
$J_{eff}$ states in a quasi one dimensional antiferromagnetic spin chain hexagonal Iridates Sr$_3$MIrO$_6$ (M=Mg, Zn, Cd): an $ab-initio$ comparative perspective}
\author {Roumita Roy}
\email{roumita19221103@iitgoa.ac.in}
\author{Sudipta Kanungo}
\email{sudipta@iitgoa.ac.in}
\affiliation{School of Physical Sciences, Indian Institute of Technology Goa, Goa-403401, India}
\date{\today}
\begin{abstract}
We employ first-principles density-functional theory, to perform a comparative investigation of the effect of the spin-orbit coupling (SOC) on the electronic and magnetic properties of three experimentally synthesized and characterized hexagonal perovskites Sr$_3$MIrO$_6$(M=Mg, Zn, Cd). The electronic structure calculations show that in all the compounds, Ir is the only magnetically active site in +4[5$d^5$] configuration, whereas M$^{+2}$ (M=Cd, Zn, Mg), remains in nonmagnetic states with Cd/Zn and Mg featuring $d^{10}$ and $d^{0}$ electronic configurations, respectively. The insulating gap could be opened by switching on the correlation parameter $U$ for Sr$_3$CdrO$_6$ and Sr$_3$ZnIrO$_6$ which qualifies it to be a correlated Mott insulator. However, in the case of Sr$_3$MgIrO$_6$ both $U$ and antiferromagnetic ordering is not enough and the gap could only be opened by including the SOC which classifies it to fall under the category of a typical SOC Mott insulator. The $j_{eff}$ states are visualized from the orbital projected band structure. The magnetism is studied from the point of view of exchange interactions and magnetocrystalline anisotropy in the presence of the SOC. We also present the comparative analysis of the renormalized impact of SOC on the three compounds, which shows that all the three compounds fall under the $intermediate$ coupling regime, where Sr$_3$MgIrO$_6$ is comparatively closer to the atomic $j_{eff}=\frac{1}{2}$ picture from the others.
\end{abstract}
 
\maketitle

\section{Introduction}
Iridates provide a fertile ground to understand the delicate interplay
amongst various energy scales that include Coulomb correlation, Hund’s
coupling, crystal-field splitting, exchange interactions, bandwidth and spin-orbit coupling (SOC). While the last decade has seen a major boom in the studies revolving around Iridates \cite{Krempa2014,Rau2016,Pedersen2016,Cao2018,Bhowal2021}, the major thrust was provided by the celebrated work on Sr$_2$IrO$_4$\cite{Kim2008,Kim2009}, where SOC Mott insulating state has been shown in the octahedral environment of Ir by the combined effect of the strong SOC and the Hubbard $U$, as a result the supposedly half-filled band split into $j_{eff}$=$\frac{1}{2}$ lower and upper Hubbard bands. This further allows us to study the magnetism of such systems in terms of the new good quantum number $j$, as derived from the atomic $j-j$ $coupling$ description
\cite{Wang2011,WanPRB2011,Jackeli2009,Moon2008,Cao2014,Bhowal2015} in the presence of the local uniform octahedral environment. 


\begin{figure*}
\begin{center}
\includegraphics[width=15cm]{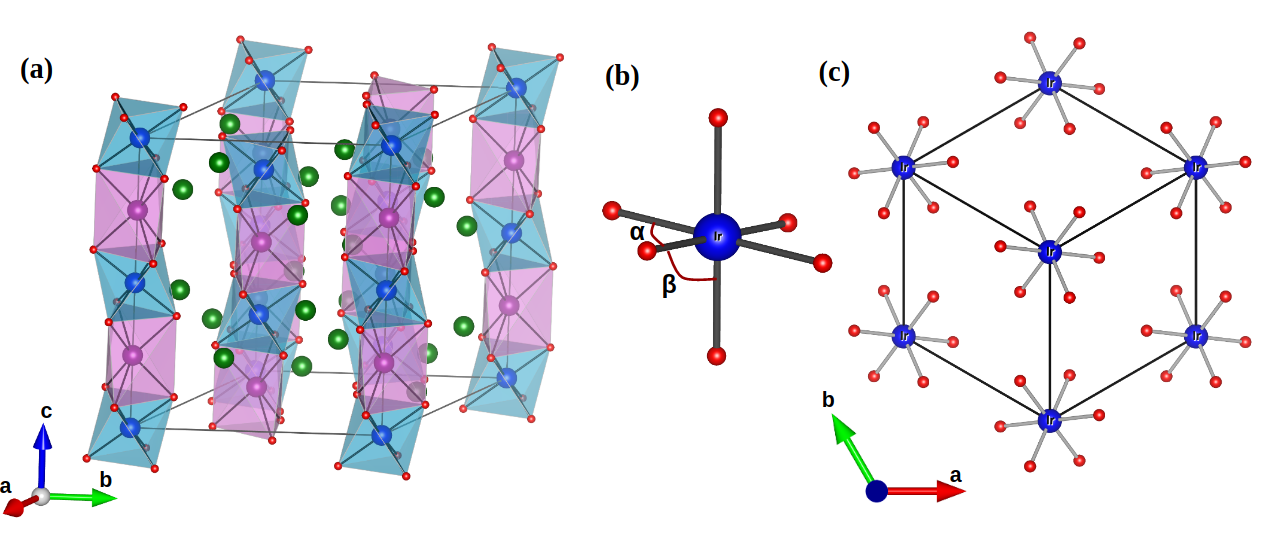}
\end{center}
\caption{ (a) Crystal structure of Sr$_3$MIrO$_6$ (M=Mg, Zn, Cd). The Sr, M, Ir and O atoms are 
represented by green. violet, blue and red spheres respectively. The IrO$_6$ octahedra and the 
MO$_6$ trigonal prism is shown in blue and violet polyhedras respectively. (b) IrO$_6$ octahedra 
(c) Hexagonal arrangement of Ir sublattice as seen from the chain direction.}
\label{fig:Fig1}
\end{figure*}

Interestingly, the common notion is that the half-filled Iridates belong to the 
strong SOC regime, where the atomic $j-j$ coupling prescription would be the most suitable 
description and the emergence of the $j_{eff}$ state is contemplated to be an obvious phenomenon. However, inside the 
solid in the non-cubic crystal field environment, the atomic SOC can be screened heavily and the 
effective renormalized strength of SOC is always not sufficient to derive the anticipated 
$j_{eff}$ states in Iridates. The pentavalent Iridates in $5d^4$ sconfiguration is expected to show 
$j_{eff}=0$ nonmagnetic state\cite{Dey2016,Ming2018} as per the atomic $j-j$ coupling 
descriptions. However, in the last few years, there have been several studies on Ba$_2$YIrO$_6$, 
Sr$_2$YIrO$_6$, Sr$_2$GdIrO$_6$\cite{Bhowal2015}, Ba$_3$ZnIr$_2$O$_9$\cite{ANag2019} 
Ba$_2$YSb$_{1-x}$Ir$_x$O$_6$ \cite{MSKhan2021}, Sr$_3$(Li/Na)IrO$_6$\cite{AB2022,arxiv2021} across several compositions of the Iridates which reveals the breakdown of the 
$j_{eff}=0$ picture. This may be because of the dominance of Hund's coupling over 
SOC\cite{Cao2018}, band structure\cite{Bhowal2015}, non-cubic crystal field\cite{MSKhan2021}, quantum spin fluctuation\cite{AB2022} or effective normalization of the SOC due to the 
spin\cite{Khomskii2014} and in all these cases the key factor is the strong competition amongst multiple energy scales in the case of Iridates.

For the tetravalent Iridates in $5d^5$ ($S=\frac{1}{2}$) state the effective SOC 
strength is comparatively higher than its pentavalent $5d^4$ ($S=1$) counterpart. This allows for the competition among 
the different scales to be even tighter between $S=\frac{1}{2}$ (atomic $L-S$ coupling) and 
$j_{eff}=\frac{1}{2}$ (atomic $j-j$ coupling) descriptions. However, even in the tetravalent 
Iridates, there are several reports on Sr$_2$CeIrO$_6$\cite{Kanungo2016}, 
Sr$_3$CuIrO$_6$\cite{Yin2013,Liu2012}, which reveals strong mixing between $j_{eff}=\frac{1}{2}$ and 
$j_{eff}=\frac{3}{2}$ in addition to the competing energy scales of the exchange interactions, 
bandwidth and mixing of \textit{t$_{2g}$}-\textit{e$_{g}$} orbitals \cite{t2geg} thus the atomic $j_{eff}=\frac{1}{2}$ state description is not completely valid. Hence the 
impact of SOC cannot be generalized within this family of Iridates and one should not decide the effect of SOC 
in deriving the $j_{eff}$ states on general grounds by looking at the nominal 
electronic valence state only.

In the present study, we have considered the hexagonal perovskites family of compounds with the 
general formula A$_3$BB$^/$O$_6$, where the A site is an alkaline earth metal, and the B and 
B$^/$ sites belong to the transition metals and are the magnetic sites. The available literature 
on this family is vast due to the tunability of A, B and B$^/$ sites. The popular choice for the 
A site includes Sr and Ca. With Sr there are examples of magnetic excitation \cite{SCuPtO,SNiIrO}, Griffiths phase like behaviour \cite{SCuRhO}, non-collinear magnetism \cite{SZnRhO} and classical spin-liquid behaviour \cite{SNiPtO}, reported in the literature. While with the latter, i.e Ca we have reports of antiferromagnetic insulator \cite{CFeRhO}, partially disordered antiferromagnetic phase \cite{CCoRhO1}, superparamagnetic clusters \cite{CCoRhO2}, multiferrocity \cite{CCoMnO} and so on.  What makes these systems even more interesting is that the effective structural dimensionality of the system is lower than three dimensions. The structure comprises alternating, face-shared BO$_6$ trigonal prism and B$^/$O$_6$ octahedra connected in a chain-like fashion along the crystallographic c-axis. The transition metal sub-lattice forms a hexagonal arrangement in the $ab$ plane and hence the nomenclature. The presence of isolated spin chains with localized magnetic moments provides an ideal ground to manifest low dimensional magnetism with a prominent signature of quantum fluctuations for small effective spin systems. Our primary focus is on systems where B$^/$ site is occupied by an Ir atom so that the investigation of $j_{eff}$ states can be realized in the strong SOC limit. While hexagonal perovskites have been discussed a lot in the context of low-dimensional spin systems, nevertheless the rise of SOC driven $j_{eff}$ states in the presence of low dimensionality is yet to be explored in detail. A few examples of such hexagonal Iridates, found in the literature include Sr$_3$CuIrO$_6$ \cite{Liu2012}, Sr$_3$CoIrO$_6$ \cite{Mikhailova2012}, Sr$_3$NiIrO$_6$ \cite{Sarkar2010,Birol2018}, Sr$_3$NaIrO$_6$ \cite{AB2022}, Sr$_3$LiIrO$_6$ \cite{Ming2018}. Point to be noted here that Sr$_3$(Na/Li)IrO$_6$\cite{Ming2018} and Sr$_3$(Ni/Co)IrO$_6$ \cite{Ou2014} are believed to depict $j_{eff}=\frac{1}{2}$ state at the Ir site, however, very recent studies\cite{AB2022,Yin2013,Liu2012,arxiv2021} show that Ir is no longer in the atomic $j-j$ $coupling$ regime. Therefore, detailed material-specific electronic structure investigation is indispensable to understand the effective impact of SOC in deriving the electronic structure. This makes our current work even more relevant where a material-specific description is portrayed that highlights the impact of SOC in the compound, from the microscopic point of view.

In the current paper, we perform a comparative study of the electronic and magnetic properties of three experimentally synthesized and characterized hexagonal Iridates, Sr$_3$CdIrO$_6$ (SCIO), Sr$_3$ZnIrO$_6$ (SZIO) and Sr$_3$MgIrO$_6$ (SMIO) \cite{Segal1996,SMgIO1997}. As per previous experiments all three compounds show antiferromagnetic ordering. The transition temperatures are reported to be 22K and 19K for SCIO and SZIO respectively \cite{Segal1996}, whereas in the case of SMIO the susceptibility vs temperature curve shows maxima at 13K, however, the exact value of T$_N$ is inconclusive \cite{SMgIO1997}. Further extensive microscopic analysis on SZIO predicts the T$_N$ to be of the order of 17K \cite{SZIO1996,McClarty2020}. In this work, we perform a relative analysis of the electronic structure and properties of these three compounds and address the pertinent question about the diversified impact of SOC on different materials belonging to the Iridate family. Our initial study includes analyzing the structural and electronic properties of these three systems which reveals that Ir in +4[5$d^5$] configuration is the only magnetically active site, whereas Cd, Zn and Mg remain inactive with an inert configuration. The point to be noted here is that in this situation we are dealing with two extremities with nonmagnetic Zn/Cd in closed shell ($d^{10}$) and Mg in open shell ($d^0$) configurations. Our study is crucial to realize the evolution of magnetism and the effect of SOC in these iso-structural and iso-electronic hexagonal Iridates by modification at only the non-magnetic sites which apparently should not modify the influence of SOC. Our study microscopically reveals the fact that SMIO falls under the category of the relativistically driven Mott insulator with large magnetocrystalline anisotropy energy, whereas SCIO and SZIO are correlation-driven Mott insulators  This is driven by the complex energy landscapes involving electronic correlation, bandwidth, crystal field splitting and SOC. In the following sections, we reveal that although the three compounds lie in the intermediate regime of complete $L-S$ and $j-j$ coupling schemes, yet the footprints of SOC are found to be very material-specific.

\begin{table*}
\centering
\begin{tabular}{| c | c | c c | c | c c c |} 
\hline
Compound  & Method & & Lattice Constant ($\AA$) & Sr & & O &  \\
   & & a & c & x & x & y & z \\ [0.8ex] 
   \hline
    & &  & & &  &  &  \\
SMIO & Expt & 9.666 & 11.103 & 0.364  & 0.174 & 0.023 & 0.113 \\
 & GGA+$U$ & - & - & 0.365 & 0.173 & 0.021 & 0.116\\
  & &  & & &  &  &  \\
  SZIO & Expt & 9.633 & 11.203 & 0.363 & 0.172 & 0.020 & 0.112 \\
 & GGA+$U$ & - & - & 0.364 & 0.174 & 0.020 & 0.113\\
  & &  & & &  &  &  \\
 SCIO & Expt & 9.657 & 11.604 & 0.362 & 0.173 & 0.018 & 0.107 \\
 & GGA+$U$ & - & - & 0.363 & 0.180 & 0.022 & 0.105\\
  & &  & & &  &  &  \\
\hline
\end{tabular}
\caption{ Experimental and theoretically optimized atomic internal coordinates for Sr and O atoms. The lattice constant was kept fixed at experimentally obtained values as reported in \cite{Segal1996,SMgIO1997}}.
\label{table:1}
\end{table*}

\section{Calculation methodology}
The DFT calculations were performed within the 
 plane-wave based basis set of 500 eV cut-off on a pseudopotential framework with Perdew-Burke-Ernzerhof 
 (PBE)\cite{Perdew1996} exchange-correlation functional as implemented in the Vienna ${ab-initio}$ simulation package (VASP)\cite{Kresse1993,Kresse1996}. The effect of electron-electron Coulomb correlations for the Ir-5$\textit{d}$ states was taken into account via onsite Hubbard $U$ ($U_{eff}$=$U$-J$_H$)\cite{Anisimov1993,Dudarev1998}. The SOC effect has been incorporated in the calculations through relativistic corrections to the original Hamiltonian \cite{Hobbs2000}. We used 5$\times5\times3$ k-mesh in the Brillouin zone (BZ) for the self-
 consistent calculations. The experimentally obtained structures were optimized by relaxing the atomic 
 positions towards equilibrium until the Hellmann-Feynman force becomes less than 0.001 eV/$\AA$, keeping 
 the lattice parameters fixed at the experimentally obtained values.

\section{Crystal Structure}
The three hexagonal Iridates under discussion have a K$_4$CdCl$_6$ type structure in the rhombohedral space group (R$\bar{3}$c). For the conventional crystal structure the Sr, M, Ir and O atoms occupy the 18$e$(x, 0, 0.25), 6$a$ (0, 0, 0.25), 6$b$ (0, 0, 0) and 36$f$ (x, y, z) Wyckoff positions respectively. Fig.\ref{fig:Fig1}(a) shows the linear 1D chain-like arrangement of the face-shared IrO$_6$ octahedra and the MO$_6$ trigonal prism, thus forming a Ir-M-Ir-M chain-like structure along the global c-direction with the Ir-M-Ir angle being  180$^o$. The point to be noted here is that the above mentioned consecutive Ir-M-Ir-M chains are not connected amongst each other, so effectively the crystal structure can be considered to consist of a collection of virtual 1D chains. The IrO$_6$ octahedras are slightly tilted towards the $a-b$ plane, such that the global z-axis doesn't coincide with the octahedral axis. The Sr atoms lie within the hollow space in between the linear chains. The hexagonal arrangement of Ir atoms can be visualized in Fig.\ref{fig:Fig1}(b).  The lattice parameters and atomic coordinates for Sr$_3$MgIrO$_6$ (SMIO), Sr$_3$ZnIrO$_6$ (SZIO) and Sr$_3$CdIrO$_6$ (SCIO)  are mentioned in Table \ref{table:1}. The lattice constant along the global c axis increases with the increase in the radius of the M atom from Mg to Cd. We find that post structural optimization the atomic positions of the M and Ir sites do not alter from the experimental Wyckoff positions. Even in the case of Sr and O atoms, the structure doesn't deviate much from the experimental case. Table \ref{table:2} lists selected bond lengths and bond angles for the three compounds. Within the IrO$_6$ octahedra, the bond lengths are equal. However, the bond angles deviate from the ideal 90$^o$, which causes the Ir-$5d$ orbitals to experience a non-cubic crystal field. This distortion is more pronounced in the case of SMIO than SZIO, followed by SCIO. This trend is consistent with the ionic radius of the M$^{2+}$ ions, which is smallest for Mg$^{2+}$, followed by Zn$^{2+}$ and Cd$^{2+}$.  

\begin{table}
\centering
\begin{tabular}{| c | c | c | c |} 
\hline
 & SMIO & SZIO & SCIO\\
 \hline
 Ir-O & 2.04 & 2.03 & 2.05\\
 M-O & 2.17 & 2.20 & 2.35\\
 O-Ir-O ($\alpha$) & 84.41 & 85.08 & 87.96\\
 O-Ir-O ($\beta$) & 95.58 & 94.92 & 92.04\\
\hline
\end{tabular}
\caption{ Bond lengths ($\AA$) and bond angles ($^o$) for theoretically optimized crystal structure for SMIO, SZIO and SCIO.}
\label{table:2}
\end{table}

\begin{figure*}
\begin{center}
\includegraphics[width=15cm]{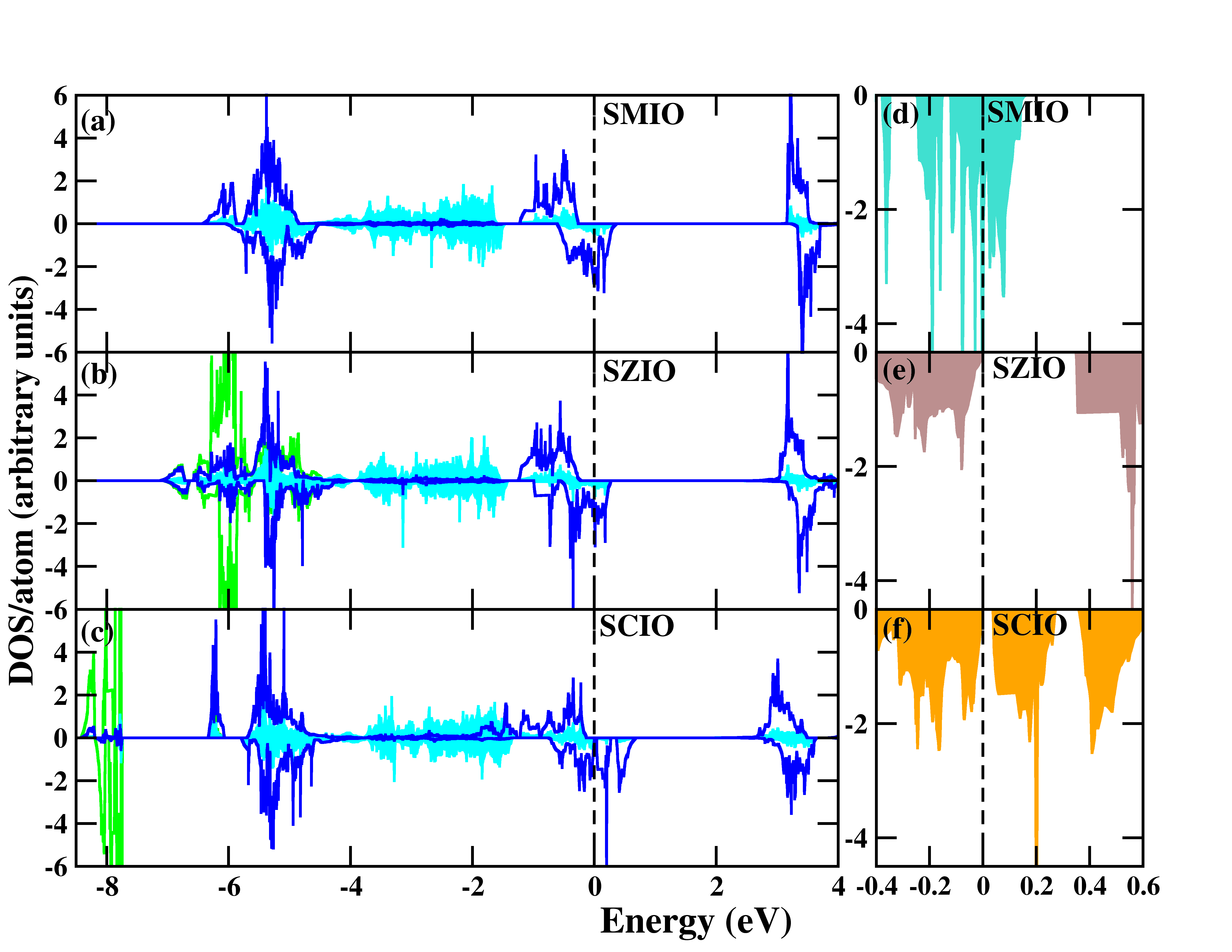}
\end{center}
\caption{  The left column represents the GGA+$U$ ($U_{eff}$=2 eV) density of states for (a) SMIO (b) SZIO and (c) SCIO. The Ir-$5d$ and O-$2p$ states are represented by blue and cyan curves respectively. The green curves represent the Zn-$3d$ and Cd-$4d$ states in (b) and (c) respectively. The density of states in the minority spin channel is shown in the right column for (d) SMIO with GGA+U ($U_{eff}$=2 eV)+AFM (e) SZIO with GGA+U ($U_{eff}$=3 eV)+FM and (f) SCIO with GGA+U ($U_{eff}$=2 eV)+FM, which represents the zoomed version of the DOS as shown in (c). The Fermi energy level is set to zero in the energy scale.}
\label{fig:Fig2}
\end{figure*}

\begin{figure}
\begin{center}
\includegraphics[width=9cm]{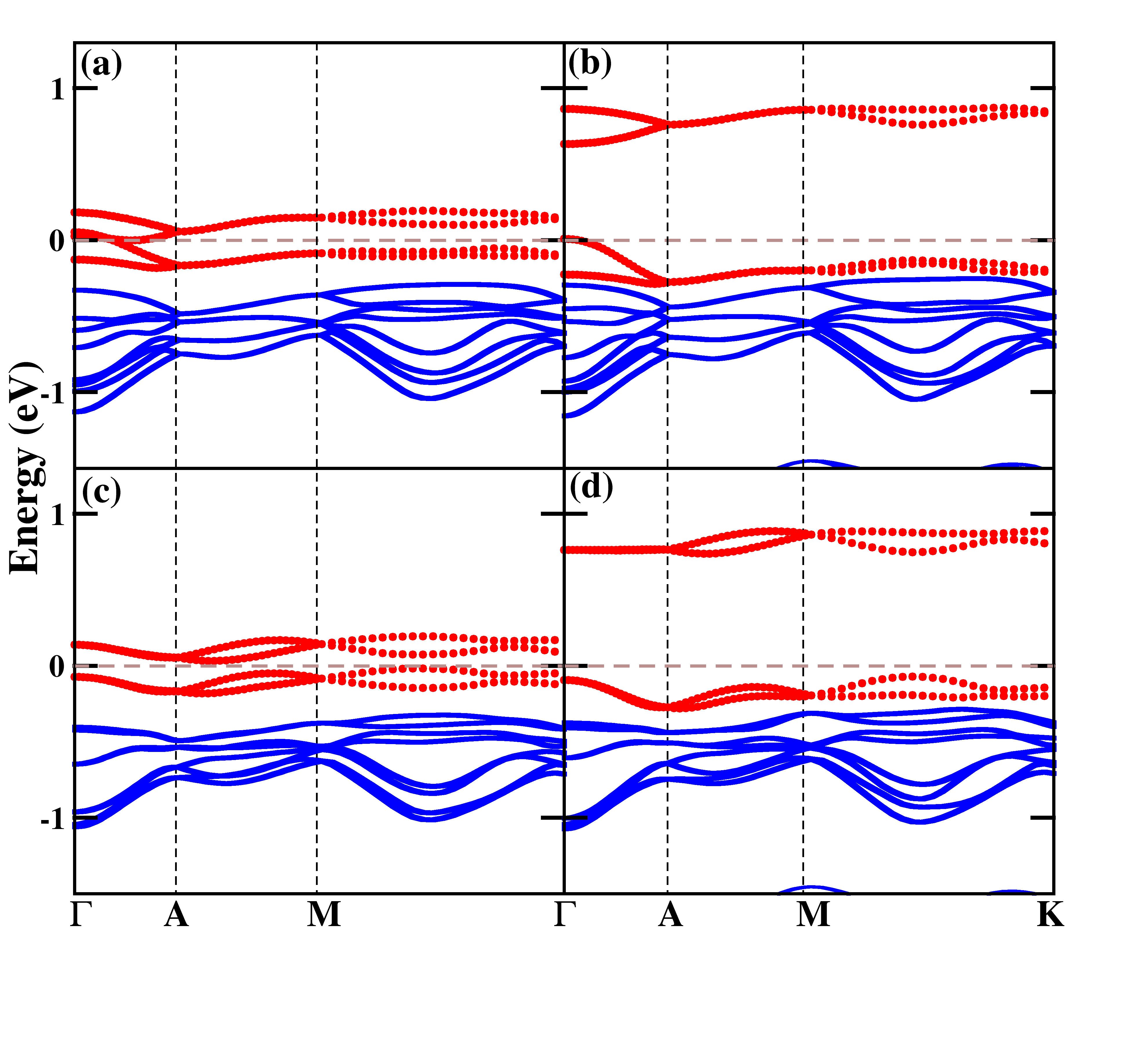}
\end{center}
\caption{Ir-$5d$ projected band structure for SMIO along the high symmetry K-points with (a) GGA+FM+SOC [001] (b) GGA+$U$+FM+SOC [001] (c) GGA+AFM+SOC [001] and (d) GGA+$U$+AFM+SOC [001]. The blue curves represent the Ir-$5d$ states and the highlighted red curves represent the j=$\frac{1}{2}$ states. The Fermi energy level is set at zero in the energy scale.}
\label{fig:Fig3}
\end{figure}

\begin{table}
\centering
\begin{tabular}{| c | c | c | c |} 
\hline
 & SMIO & SZIO & SCIO\\
 \hline
 Sr &  0.00 & 0.00 & 0.00\\
 M & 0.00 & 0.01 & 0.02\\
 Ir & 0.61 & 0.63 & 0.75\\
 O & 0.06 & 0.06 & 0.03\\
\hline
\end{tabular}
\caption{ The calculated value of spin magnetic moment (in $\mu_B$/site ) for GGA+$U$ ($U_{eff}$=2 eV) in SMIO (M=Mg, Zn and Cd).}
\label{table:3}
\end{table}

\section{Electronic Structure}
Fig.\ref{fig:Fig2}(a-c) shows the GGA+$U$ ($U_{eff}$=2 eV) orbital projected DOS for all the three compounds.  In all three cases, the Sr dominated bands lie far away from the Fermi energy and are not shown within the energy range. Sr is in an inert state with a valency of +2, with no major contribution to the DOS at E$_f$.  In the case of SCIO and SZIO, the Cd-$4d$ and Zn-$3d$ states respectively are completely filled in both the spin channels. For SZIO, the Zn states lie 6 eV below the E$_f$ whereas for SCIO the Cd states lie even farther i.e $\approx$ 8 eV below the E$_f$. This is in conformity with Zn and Cd being in +2 valence states. On a similar footing Mg in SMIO is in +2 valence state with completely empty $d$ orbitals. The DOS lying between -6 to -4 eV arises from the strong hybridization between the Ir-$5d$ and O-$2p$ states, as reflected from Fig.\ref{fig:Fig2}(a-c). The O-$2p$ DOS is mainly concentrated within the energy range -4 to -1 eV and is well separated from the Ir-$5d$ states near the Fermi energy, in the case of SZIO and SMIO. In the octahedral environment, the Ir-$5d$ states split into \textit{t$_{2g}$} and \textit{e$_{g}$} states, with the \textit{e$_{g}$} states being completely empty in both the spin channels and can be seen in Fig.\ref{fig:Fig2}(a-c) in the energy range of 3 to 4 eV. The \textit{t$_{2g}$} states are completely filled in the majority spin channel and partially filled in the minority spin channel. The values of the spin magnetic moment are listed in Table \ref{table:3}. We find that the moment at the Ir site increases as we move from SMIO to SCIO. The net moment for all three systems is found to be 1 $\mu_B$ per formula unit. Looking at the combined results of DOS and the magnetic moment we conclude that Ir is in +4[$5d^4$] with a low spin state of S=$\frac{1}{2}$ in all three compounds. The absence of substantial value of moment at the O site, further suggests that the magnetic moment at the Ir site is quite localized. Thus this system can be considered to be an arrangement of spin $\frac{1}{2}$ linear chains running along the c-direction. The interactions within this chain and with its neighboring ones are discussed in subsequent sections. Furthermore, as compared to other Iridates \cite{Kim2008}, the bandwidth of the \textit{t$_{2g}$} states here is much narrower due to reduction in the electronic hopping as a result of lower structural connectivity. Thus it is natural to expect that it would be possible to open up the insulating gap with the inclusion of a reasonable value of on-site Hubbard $U$. Counterintuitively, the insulating nature is only possible with $U$ value of 2 eV for the SCIO as seen in Fig.\ref{fig:Fig2}(c) and (f). On the other hand for SZIO, a larger value of $U$=3 eV is required to open up the gap as evident in Fig.\ref{fig:Fig2}(b) and (e). We believe that the increased structural distortion in SZIO, as compared to SCIO, calls for a larger value of $U$ in SZIO, to open up this gap. While a marginal gap opens up at the Fermi energy for SCIO and SZIO for $U$ values of 2 eV and 3 eV respectively, SMIO essentially retains its metallic character with a large $U$ value even up to 4 eV at the Ir site.  Furthermore it is widely known that on imposing the AFM order, it is possible to open up a band due to a reduction in bandwidth. Nevertheless, even with the introduction of antiferromagnetic ordering, SMIO holds its metallic state in the spin-down channel as can be seen in Fig.\ref{fig:Fig2}(d).

Fig. \ref{fig:Fig3}(a) and (b) show the GGA+FM+SOC and GGA+$U$+FM+SOC bandstructures for SMIO respectively whereas Fig.\ref{fig:Fig3}(c) and (d) depicts the GGA+AFM+SOC and GGA+$U$+AFM+SOC bandstructures, where the ground state AFM ordering has been considered. The twelve bands near the Fermi energy arise from the \textit{t$_{2g}$} states of the two Ir sites in the primitive lattice. In Fig.\ref{fig:Fig3}(a), the bands overlap at the Fermi energy level near the high symmetry $\Gamma$ point. However, as soon as we switch on the electronic correlation ($U$), a small gap is introduced as shown in Fig.\ref{fig:Fig3}(b). The inclusion of the ground state antiferromagnetic ordering enhances the insulating gap in SMIO. A notable point here is that even in the absence of $U$, an insulating gap is obtained with AFM configuration as seen in Fig.\ref{fig:Fig3}(c). This is majorly due to the reduction of bandwidth in SMIO, driven by AFM exchange.  Similar results were previously obtained for isostructural Sr$_4$IrO$_6$ \cite{Ming2018} and  Ca$_4$IrO$_6$ \cite{Calder2014} where SOC was essential to introduce the insulating state. SOC also has a significant influence on the band dispersion of the Ir-$5d$ states for SMIO, the \textit{t$_{2g}$} states separate into the $j_{eff}$=$\frac{1}{2}$ doublet and the $j_{eff}$=$\frac{3}{2}$ quartet. The 8 bands arising out of the latter lie in the energy range -0.5 to -1 eV and are completely occupied. The remaining unpaired electrons from the Ir atom goes to the $j_{eff}$=$\frac{1}{2}$ doublet and the degeneracy breaks due to partial occupancy. We can visualize the overlapping (Fig.\ref{fig:Fig3}(a)) and the well separated (Fig.\ref{fig:Fig3}(c)) $j_{eff}$=$\frac{1}{2}$ states near the Fermi energy in SMIO with the FM and ground state AFM order respectively. The inclusion of Hubbard correlation $U$ in Fig.\ref{fig:Fig3}(b and d) further separates the $j_{eff}$=$\frac{1}{2}$ states giving rise to a completely filled $j_{eff}$=$\frac{1}{2}$ lower Hubbard bands and completely empty $j_{eff}$=$\frac{1}{2}$ upper Hubbard bands with a gap of the order of 0.6 eV and 0.8 eV respectively. At this point we need to emphasize the fact that even in the absence of Coulomb interaction $U$, SOC transforms SMIO from an AFM metallic to an AFM insulating state. This further denotes the supremacy of SOC interactions which reduces SMIO to a half-filled $j_{eff}$=$\frac{1}{2}$ spin-orbit coupled Mott insulator. Here the size of the gap is comparable to the gap size in iso-structural Sr$_4$IrO$_6$ \cite{Ming2018}, but much larger as compared to Sr$_2$IrO$_4$ \cite{Kim2008}. This signifies the importance of the presence of isolated Ir octahedras, which reduces the electronic correlation and enhances the SOC. The point to be noted here is that in Sr$_4$IrO$_6$, Ir is in an ideal cubic crystal field, thus intuitively the effect of SOC should be stronger in comparison to SMIO where Ir occupies a distorted octahedra. However, we find the effect of SOC is similar in both cases, which establishes that SOC indeed is the most crucial interaction in SMIO.

\begin{figure*}
\begin{center}
\includegraphics[width=18 cm]{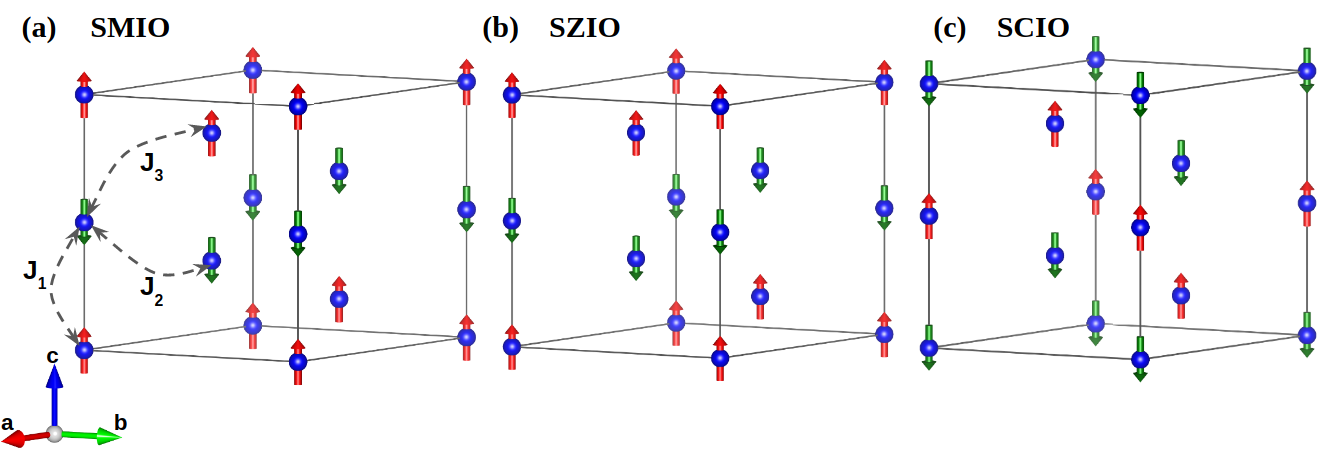}
\end{center}
\caption{ The magnetic ground state obtained with GGA+U+SOC for (a) SMIO (b) SZIO and (c) SCIO. The blue spheres represent the Ir atoms. The red and green arrows represent up and down spins respectively. The exchange interactions J$_1$, J$_2$ and J$_3$ are marked in (a).}
\label{fig:Fig4}
\end{figure*}

\section{Magnetism}
In this section, we discuss the magnetic ground state of the three systems. Previous studies suggest that the AFM ordering temperature in the case of hexagonal Iridates is relatively low as compared to other Iridate systems due to reduced connectivity amongst Ir ions\cite{Ming2018}. The reduced structural connectivity amongst the Ir-M-Ir-M chains which host the magnetic Ir site, further reduces the electron hopping integral resulting in limited magnetic exchange.  As a starting point, we consider the magnetic unit cell to be the same as the crystallographic hexagonal unit cell and take into account various possible spin configurations to obtain magnetic ground states for SMIO, SZIO and SCIO as shown in Fig.\ref{fig:Fig4}(a-c). Point to be noted here is that in the absence of SOC, we find SMIO and SZIO to be FM in nature. The true AFM ground state in these two compounds could only be realized with the inclusion of SOC, which further indicates the importance of SOC interactions even in deriving the correct magnetic ground state. Our calculations reveal that the spins prefer to orient along the chain axis, i.e. the global $z$ direction in an antiferromagnetic fashion. 

\begin{table*}
\centering
\begin{tabular}{| c | c c | c c | c c |} 
 \hline
 System &  J$_1$ & (intra-chain) & J$_2$ & (NN inter-chain) & J$_3$  & (NNN inter-chain)\\
  & Ir-Ir distance  & value(type) & Ir-Ir distance & value(type)  & Ir-Ir distance & value(type)\\ [0.8ex] 
 \hline
 SMIO & 5.55 & 5.784(AFM) & 5.88 & 0.213(FM) & 6.70 & 0.014(AFM)\\
 SZIO & 5.60 & 17.576(AFM) & 5.87 & 0.391(FM) & 6.70 & 0.008 (AFM)\\
  SCIO & 5.80 & 16.303(AFM) & 5.90 & 0.827(AFM) & 6.78 & 0.008(AFM)\\
\hline
\end{tabular}
\caption{ The calculated magnetic exchange interactions for different paths for SMIO, SZIO and SCIO, as shown in Fig.\ref{fig:Fig4}(a). The Ir-Ir distance is mentioned in $\AA$ and the values of the magnetic exchange interaction in meV.}
\label{table:4}
\end{table*}

Further to analyze the nature of the magnetic exchange, we compute the exchange interaction energies. This is implemented by mapping the DFT total energies of several artificially constructed spin configurations into the Heisenberg Hamiltonian \cite{Helberg1999,Mazurenko2006,Xiang2011} of the form of, $E^{Tot}$=$\sum_{ij}J_{ij}S_{i}S_{j}$, where J$_{ij}$ is the magnetic exchange interaction between the i$^{th}$ and j$^{th}$ sites and $S_{i}$ and $S_{j}$ are the effective spins at the corresponding sites. There are a few drawbacks of this method including the choice of correct spin configurations, exchange path, and exchange-correlation functional. Nevertheless, it is known to provide an estimate of the strength and nature of the magnetic exchange interactions which is much required for qualitative understanding of the magnetic properties of the various classes of materials \cite{Sarkar2010,Kanungo2014,Sannigrahi2019,RRoy}. Based on the chain-like structure of the systems, we have considered three independent
possible exchange interaction pathways as can be seen in Fig.\ref{fig:Fig4}(a). Amongst them, J$_1$ represents the Ir-Ir intra-chain interaction, whereas J$_2$ and J$_3$ are the inter-chain Ir-Ir interactions amongst the adjacent chains. J$_2$ and J$_3$ take into account the nearest-neighbor (NN) and next-nearest-neighbor (NNN) inter-chain interactions respectively. The values of the Ir-Ir bond length associated with the various J's is listed in Table \ref{table:4}. The values and nature of the magnetic exchange interaction (J's) considering all the spins to be pointing along the z-direction with a spin value of S=$\frac{1}{2}$, is listed in  Table \ref{table:4} for SMIO, SZIO and SCIO. 

We find that the intra-chain interaction J$_1$ is uniformly the strongest among all the interactions for SMIO, SCIO as well as SZIO. It is anti-ferromagnetic in nature which is crucial in establishing the overall AFM ground state. The point to be noted here is that J$_1$ is of the same order for SCIO and SZIO, but is almost three times smaller for the case of SMIO. One probable reason could be the presence of $d$-electrons in the closed shell configuration of Cd$^{2+}$ in SCIO and Zn$^{2+}$ in SZIO which aids the electron-hopping. This is however not the scenario in SMIO where the non-magnetic Mg$^{2+}$ ion is in open shell configuration. A similar incident has also been previously reported in the case of rock-salt double-perovskite (Sr$_2$BOsO$_6$; B= Sc, Y, In) \cite{2Kanungo2016}, where the shell configuration of the non-magnetic site dictates the strength of the magnetic exchange interactions of the 5$d$ elements. The nearest neighbor inter-chain interaction is FM for the case of SMIO and SZIO and AFM for the case of SCIO. However, the strength of J$_2$ is two orders of magnitude lower compared to J$_1$. On the other hand, the next nearest neighbor interaction is almost negligible for all three systems. These findings are crucial in establishing the fact that although the compounds are structurally three-dimensional, nevertheless from the point of view of magnetic interactions it reduces to an arrangement of spin $\frac{1}{2}$ chains along the z-direction. Thus effectively they are quasi one-dimensional in nature with the absence of any interactions amongst the consecutive chains. Therefore it serves as one of the exemplary systems to study the physics of 1D spin chains in the presence of SOC in a $j_{eff}$ basis. Furthermore, from mean field calculations, we find that the transition temperature for SCIO is of the same order as that of SZIO. For SMIO, our calculations reveal that the transition will occur at a much lower temperature which is expected to be of the order of $\frac{1}{3}$rd value as that of SZIO. These results are consistent with previous experimental findings \cite{McClarty2020,SZIO1996,Segal1996,SMgIO1997}.

\begin{table}
\centering
\begin{tabular}{| c | c | c | c | c |} 
\hline
 & (m$_z$) & (o$_z$) & $\frac{o_z}{m_z}$ & E$_{MCA}$\\
 &  ($\mu_B$/site) & ($\mu_B$/site) &  & (meV/f.u)\\
 \hline
  SMIO & 0.31 & 0.52 & 1.67 & 30.44\\
 SZIO & 0.33 & 0.55 & 1.55 & 4.48\\
 SCIO &  0.37 & 0.52 & 1.40 & 2.66\\
\hline
\end{tabular}
\caption{ The calculated values of spin magnetic moment (m$_z$) and orbital magnetic moment (o$_z$) and their ratios are mentioned in the first three columns. The calculations were performed under the GGA+$U$+SOC scheme along [001] chain direction. The last column represents the calculated values of magnetocrystalline anisotropy energy for the respective AFM ground states for SMIO, SZIO and SCIO. The E$_{MCA}$ is calculated as $\lvert$ E$_{||}$-E$_{\perp}$ $\lvert$, where E$_{||}$ and E$_{\perp}$ represents the DFT total energy for the spin configuration parallel and perpendicular chain direction respectively. }
\label{table:5}
\end{table}

\section{Effect of SOC}

By now it has been established that SOC is an important energy scale for the compounds under investigation. Hence a detailed study on its effect becomes inevitable. We thus perform GGA+$U$+SOC calculations in detail to understand its underlying effects. From Table \ref{table:3} and \ref{table:5}, we can see that with the inclusion of SOC, the spin magnetic moment decreases from 0.75 to 0.37 $\mu_B$ for SCIO, 0.63 to 0.33 $\mu_B$ for SZIO and 0.61 to 0.31 $\mu_B$ for SMIO, and a pronounced orbital magnetic moment occurs at the Ir site with a value of $\sim$ 0.5 $\mu_B$/site. A large value of orbital magnetic moment in comparison to its spin counterpart further provides evidence that these systems lie in the strong SOC limit. The $\mu_{eff}$ as reported from experimental studies are 1.63, 1.71 and 1.41 for SCIO, SZIO and SMIO respectively \cite{Segal1996,SMgIO1997}. The point to be noted here is that these values deviate from the ideally expected value of the spin-only magnetic moment of 1.73, with the deviation being much more pronounced in SMIO. The reason becomes prominent from the values of the magnetic moments as obtained from our DFT results which point out a significant transfer of moment from the spin to the orbital counterpart in the case of SMIO. The ratio of the orbital magnetic moment (o$_z$) and the spin magnetic moment (m$_z$), $\frac{o_z}{m_z}$, is found to be largest for SMIO, followed by SZIO and SCIO. Thus providing a quantitative analysis of the resultant influence of SOC on these three systems. The $\frac{o_z}{m_z}$ is very high for SMIO ($\sim$ 1.67), close to the ideally expected value of 2 which is known to occur for strong  $j_{eff}$=$\frac{1}{2}$ systems \cite{Kim2008}. Nevertheless, due to reduced structural connectivity of the IrO$_6$ octahedras,  hexagonal Iridates are known to deviate from this ideal behaviour\cite{Ou2014,Ming2018}. 

The crystal structure of the systems under investigation is highly anisotropic, which further translates to the electronic and magnetic interactions. Hence a high value of magnetocrystalline anisotropy (MCA) energy is intuitive and has also been reported in literature \cite{Mikhailova2012,Sarkar2010,Wu2005}. In order to estimate the MCA energy and predict the easy axis, we compared the energies along different spin quantization axes, viz. [001] and [110] for the ground-state AFM configuration. Here [001] represents the chain direction and [110] represents the plane perpendicular to the chain direction.  We find that in all three cases, the easy axis is towards the crystallographic chain direction, which is expected in a 1D chain-like system \cite{Lefrancois2014}. We also counter-checked the total energy with a canted spin orientation for all three systems, nevertheless, the spins prefer to orient along the [001] direction. The MCA energies, computed as the energy difference between the easy and the hard axis, which in these three cases are found to be the parallel (E$_{||}$) and perpendicular (E$_{\perp}$) to the chain direction respectively, are listed in Table \ref{table:5}. From our calculations, we infer that the MCA is highest for SMIO with a substantial value of 30.44 meV/f.u. The MCA for SCIO and SZIO are comparable in magnitude and an order of magnitude lower than SMIO. This further highlights the enhanced impact of SOC in SMIO, as compared to SCIO and SZIO.

\section{Discussion and Conclusion}
\begin{figure}
\begin{center}
\includegraphics[width=9 cm]{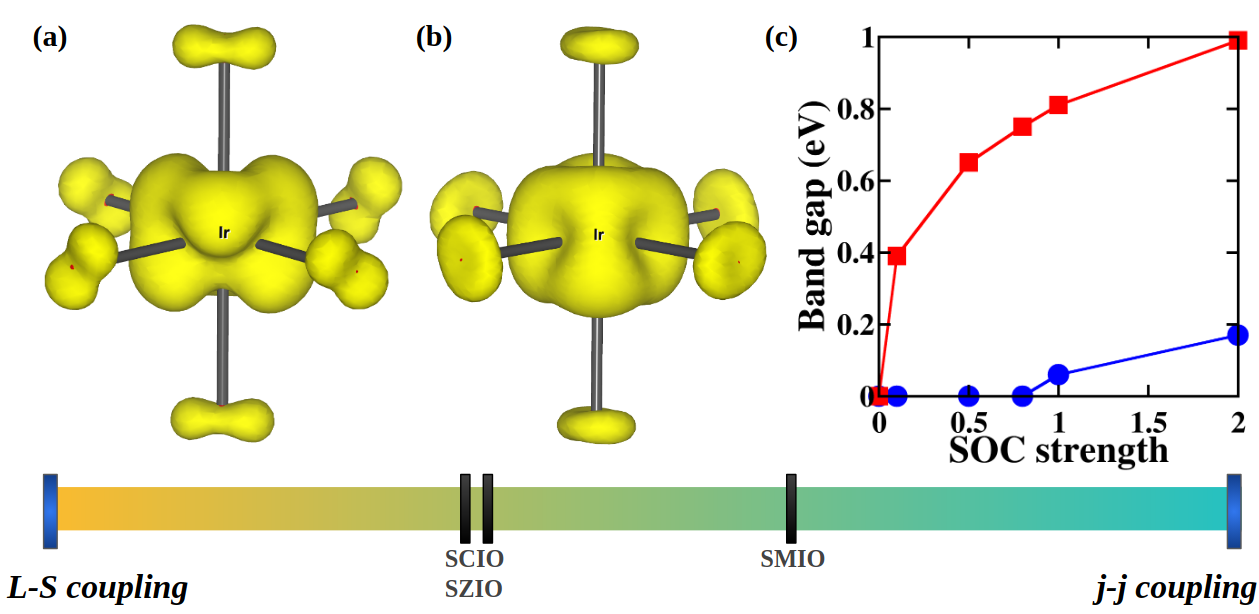}
\end{center}
\caption{Density profile of the $t_{2g}$ hole in  (a)SZIO and (b)SMIO. (c) The variation of band gap for SMIO due to tuning of SOC strength. The blue and red curves represent the band gaps (in eV) for GGA+AFM+SOC[001] and GGA+U+AFM+SOC[001] respectively.}
\label{fig:Fig5}
\end{figure}

The electronic structure calculation exhibits that SCIO and SZIO with the nonmagnetic site (Cd and Zn respectively) in closed shell configuration, fall under the category of typical correlation driven Mott insulator. On the other hand, SMIO, where the nonmagnetic site (Mg) is in an open shell electronic configuration, is a SOC driven Mott insulator. 
To further shed light on the relative influence of SOC, we look into the density profile of the $t_{2g}$ hole in SZIO and SMIO as shown in Fig. \ref{fig:Fig5}(a) and (b) respectively. The density profile can be visualized in terms of the magnetization density or the spin density, which represents the shape of the outermost partially occupied orbital. In this case, it reciprocates the $t_{2g}$ hole for the low spin state Ir-5$d^5$ configuration for both SZIO and SMIO. However from Fig. \ref{fig:Fig5}(a) and (b) the density profiles can be observed to be significantly different in both these compounds. For SMIO, the shape of the density hole is closer to what is expected for the ideal $j_{eff}$=$\frac{1}{2}$ case \cite{Rau2016,Jackeli2009,Kanungo2016}. In SZIO the spin density is distorted and is more likely to be in an intermediate picture between complete $\textit{L-S coupling}$ and complete $\textit{j-j coupling}$ \cite{Kanungo2016} scenarios. This further supports our claim that the effect of SOC is stronger in SMIO with a $d^0$ configuration as compared to SZIO (or SCIO) with  $d^{10}$ configuration. Fig. \ref{fig:Fig5} (c) represents the variation of the band gap in SMIO, with the modification of the SOC strength. The point to be noted here is that in the presence of Hubbard $U$, the insulating gap opens up for the SOC strength as low as $\frac{1}{10}^{th}$ of the intrinsic value. Thus for SMIO, the effective SOC strength is higher than the rest and competes with the electronic correlation $U$.

The strong interplay amongst electronic correlation, bandwidth, crystal field splitting, exchange interactions, structural distortion, local geometry, hybridization, and SOC is crucial in understanding the underlying electronic structure and properties of the system. The competition results in the renormalization of the associated energy scales in Iridates. This further causes the screening of strong atomic SOC effect thus resulting in the breakdown of the atomic $j-j$ coupling picture. The lattice distortions within the series are quite marginal which rules it out to be a key ingredient in driving the insulating mechanism. Rather we believe that the differentiation of SCIO, SZIO and SMIO where the first two fall under the category of correlation driven Mott insulator and the latter as a SOC Mott insulator is driven by the comparative energetics of relevant energy scales. In the context of our work, the correlation effects are more predominant in SCIO and SZIO, where the changes in DOS are significant with the variation of $U$ (See Fig.\ref{fig:FigA} in APPENDIX A). This could be attributed to the fact that unlike SMIO, the non-magnetic site in SCIO and SZIO consists of 3$d$/4$d$ transition metal ion which accentuates the correlation effects.  Another notable point is that the $t_{2g}$ - $e_g$ crystal field splitting energy increases from SCIO to SZIO to SMIO. From band structure calculations we obtain that the t$_{2g}$ bandwidth for SCIO and SZIO is $\approx$0.9 eV and $\approx$0.8 eV in SMIO. Then again for e$_g$ the bandwidth is sufficiently large for SCIO ($\approx$1.7 eV) and SZIO ($\approx$1.6 eV) as compared to SMIO ($\approx$0.9 eV). The combined effects of reduced electronic correlation, large crystal field splitting and small bandwidth in SMIO, reduce the electronic hopping and push the SMIO to behave closer to that of the atomic $j-j$ like description as
compared to the other two compounds\cite{Bhowal2015}. This can also be traced from the calculated values of magnetic exchange interaction (J’s) which are much smaller in SMIO as compared to SCIO and SZIO. The above claims have also been supported by the comparative ratio of the orbital and spin magnetic moment ($\frac{o_z}{m_z}$), which is highest for the SMIO followed by the SZIO and SCIO. All the above conditions support that the strength of effective SOC is much more pronounced in the case of SMIO than the SZIO and SCIO, however in all three cases the ideal atomic $j-j$ picture is not a proper description. Rather, they belong to the $intermideate$ coupling regime, where SMIO situating closer to the $j_{eff}$ picture as shown in the schematic diagram in Fig. \ref{fig:Fig5}.

To conclude, using first-principles DFT calculations, we have investigated the electronic structure of three compounds, Sr$_3$MgIrO$_6$, Sr$_3$ZnIrO$_6$, and Sr$_3$CdIrO$_6$.  Our study reveals that although these systems are iso-electronic, iso-structural, yet due to the combined influence of crystal structure and crystal field effects, we can differentiate them based on the impact of SOC which is found to be most crucial for the case of Sr$_3$MgIrO$_6$. The evaluated magnetic exchange interactions establish these Iridates to be magnetically low dimensional, more precisely quasi 1D in nature.  Furthermore, the magnetocrystalline anisotropy energies were evaluated and a large anisotropy is reported for Sr$_3$MgIrO$_6$. Our major findings from the comparative study on the three compounds lead us to believe that technically neither of them are in the ideal $j-j$ coupling regime, however SMIO is the closest to the ideal atomic $j-j$ picture. We hope our theoretical results will stimulate further experimental investigations on these systems. 

\section{Acknowledgements} We would like to thank Indra Dasgupta for fruitful discussions. RR thanks IIT Goa for providing research fellowship.   

\section{Appendix A}
\begin{figure}
\begin{center}

\includegraphics[width=9cm]{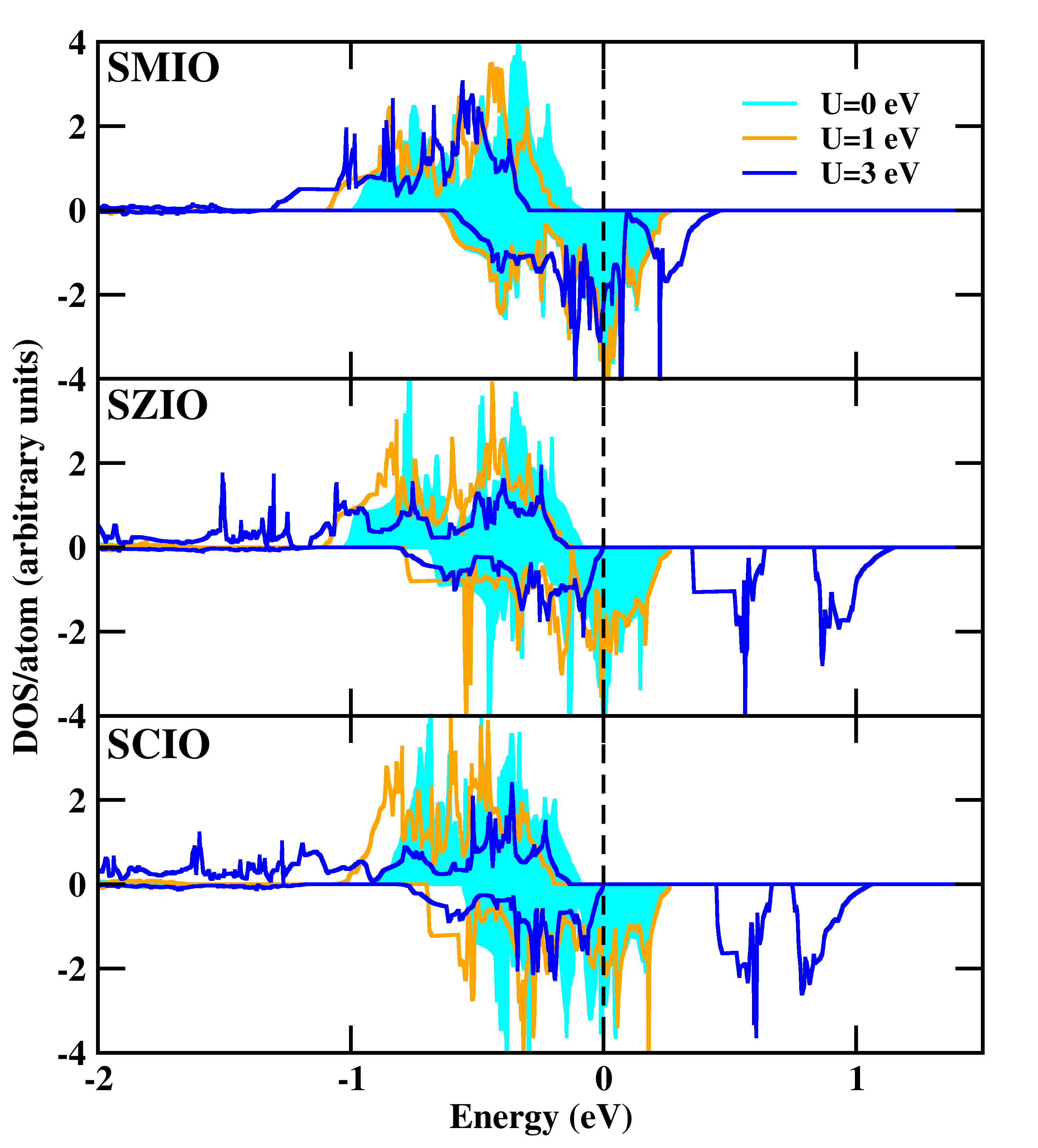}
\end{center}
\caption{The calculated GGA+$U$ Ir-5$d $ density of states with the variation of $U_{eff}$ from 0 to 3 eV is shown for SMIO, SZIO and SCIO in the top, middle and bottom rows respectively.}
\label{fig:FigA}
\end{figure}

 In the three hexagonal perovskites under discussion,  the non-magnetic sites consist of elements spanning various groups of the periodic table. Although uncanny, previously it has been seen that the non-magnetic site could also impact the electronic and magnetic properties of the system. The significant difference amongst the non-magnetic sites is the presence of 3$d$/4$d$ transition metal atoms in SZIO and SCIO. Since in transition metals the electron-electron correlation is sizeable, we study its influence on the hexagonal Iridates by tuning the value of the Hubbard $U$ parameter. The evolution of the density of states thus obtained is shown in Fig. \ref{fig:FigA}.

\end{document}